\documentclass[%
 reprint,
superscriptaddress,
 amsmath,amssymb,
 aps,prd,
]{revtex4-2}
\usepackage{amsmath}
\usepackage[section]{placeins}
\usepackage[utf8]{inputenc}
\usepackage[english]{babel}
\usepackage{float}
\usepackage{isotope}

\usepackage[dvipsnames]{xcolor}

\usepackage{soul}
\usepackage{graphicx}% Include figure files
\usepackage{dcolumn}% Align table columns on decimal point
\usepackage{bm}% bold math

\usepackage{graphicx}% Include figure files
\usepackage{dcolumn}% Align table columns on decimal point
\usepackage{bm}% bold math
\usepackage[draft, inline, nomargin]{fixme}
\fxsetup{theme=color, mode=multiuser}
\FXRegisterAuthor{bc}{1}{\color{red}BC}
\FXRegisterAuthor{ag}{2}{\color{blue}AG}
\FXRegisterAuthor{ph}{3}{\color{magenta}PH}
\newcommand{\gaggt}{Ce:GAGG}
\begin{document}

%\preprint{APS/123-QED}

\title{Testing the gallium anomaly}

\author{Patrick Huber}
\email{pahuber@vt.edu}
\affiliation{Center for Neutrino Physics, Physics Department, Virginia Tech, Blacksburg, VA 24061}

\date{\today}% 

\begin{abstract}
We study the online detection of gallium capture of mono-energetic neutrinos produced by a $^{51}$Cr radioactive source in a scintillation experiment. We find that cerium-doped gadolinium aluminum gallium garnet (\gaggt)  is a suitable scintillator which contains about 21\% of gallium per weight and has a high mass density and light yield. Combined with a highly efficient light detection system this allows tagging of the subsequent germanium decay and thus a clean distinction of gallium capture and elastic neutrino electron scattering events. With 1.5 tons of scintillator and 10 source runs of 3.4\,MCi, each, we obtain about 1700 gallium capture events with a purity of 90\% and 680,000 neutrino electron scattering events, where the latter provide a precise normalization independent of any nuclear physics. We include a detailed discussion of backgrounds and find that this configuration would allow to test the gallium anomaly at more than $5\,\sigma$ in an independent way.  

\end{abstract}

%\keywords{Suggested keywords}%Use showkeys class option if keyword
                              %display desired
\maketitle

\section{\label{sec:introduction}Introduction}

The so-called gallium anomaly is the long-standing experimental observation of a 20\% deficit of event rate in the following semi-leptonic charged current reaction
\begin{equation}
\label{eq:capture}
    \nu_e + {^{71}}\textrm{Ga} \rightarrow e^- + {^{71} }\textrm{Ge}
\end{equation}
induced by mono-energetic neutrinos with an energy of approximately $750\,\mathrm{keV}$ from the electron-capture decay of artificially produced $^{51}$Cr. This type of experiment has been conducted four times independently by the SAGE~\cite{Abdurashitov:1998ne}, Gallex~\cite{Hampel:1997fc,Kaether:2010ag} and BEST~\cite{Barinov:2021asz} collaborations; in each case a deficit of about 20\% relative to prediction was found, however the error bars  previously were too large to draw any firm conclusion.  A similar result was obtained by SAGE using a $^{37}$Ar source~\cite{Abdurashitov:2005tb}. This changed with the advent of BEST measurement which has improved the precision to a level where this result now constitutes a more than 5$\,\sigma$ statistical significance deviation from the Standard Model expectation~\cite{Berryman:2021yan}. A recent reevaluation of the underlying cross section seems to exclude this as a sole source for the discrepancy~\cite{Haxton:2023ppq}. All those experiments have in common that they are employing radio-chemical detection, whereby the reaction product ${^{71} }\textrm{Ge}$ 
is chemically extracted from the gallium and the decay back into ${^{71} }\textrm{Ga}$ is observed, effectively counting individual germanium atoms. Broadly one or more of the following reasons can be the behind this intriguing result, see also Ref.~\cite{Brdar:2023cms}
\begin{enumerate}
    \item The source strength determination is incorrect.
    \item The prediction of the interaction cross section is incorrect.
    \item The determination of the radio-chemical detection efficiency is incorrect.
    \item There is new physics.
\end{enumerate}
Since four (or five, including the $^{37}$Ar result) independent measurements based on the same technology have yielded a common result, we will investigate here the possibility of using a different experimental approach to study the very same reaction. We would also like to stress that items 1--3 have been studied in detail and with great care by the various collaborations and it seems difficult to imagine that any one of these could account for a 20\% effect.

Specifically, we study the online detection of the capture reaction in Eq.~\ref{eq:capture} by using scintillation detectors where a suitable scintillator is loaded with gallium and thus provides both the target as well as the detection medium. Capture of low-energy electron neutrinos and their measurement via scintillation has been first proposed by R. Raghavan for the LENS experiment to study solar pp neutrinos~\cite{Raghavan:1976yc}. The use of a $^{51}$Cr source in conjunction with LENS~\cite{Grieb:2006mp} using neutrino capture on indium as well as with Borexino to study elastic electron neutrino scattering~\cite{Borexino:2013xxa} have been proposed to study sterile neutrino oscillation. In essence we propose to merge both ideas and to replace indium with gallium. Since gallium is not radioactive, unlike indium, the electron scattering signal is  available at the same time as the gallium capture signal.
 In an online experiment the signature of the capture is a mono-energetic electron with an energy $E_c=E_\nu-Q_c\simeq 510\,$keV, where $Q_c=233\,$keV is the Q-value of the capture reaction. At the same time, the following reaction with the electrons in the detector will take place
\begin{equation}
\label{eq:elastic}
    \nu_e + e^- \rightarrow \nu_e + e^- \,.
\end{equation}
Elastic electron neutrino scattering is a purely leptonic process including both charged and neutral current contributions, which in the Standard Model can be precisely (at the $10^{-4}$ level) predicted~\cite{Tomalak:2019ibg,Hill:2019xqk}. Thus, if this reaction is observed at the predicted rate, it would eliminate reason 1, the source strength, as the culprit and would allow in the comparison of the rate of the capture reaction to measure the capture cross section, eliminating reason 2. Experimentally, this reaction can be distinguished, by the energy of the resulting electron, from the Ga-capture process. Online detection directly allows to address reason 3, i.e., if the result is an artefact of the radio-chemical method no deficit would show up in an online experiment. Finally, if there is indeed new physics, it will depend on the specific model for new physics, whether both rates for capture and electron scattering are modified or not. In the case of a sterile neutrino oscillation~\cite{Giunti:2010zu,Giunti:2022btk} both rates would show the same oscillatory dependence as a function of distance.  In an scintillation experiment the baseline can be measured precisely on an event-by-event basis and thus the typical oscillatory distance dependence could be observed directly at a cm-scale (as we will illustrate later) directly probing mass squared differences of up to $\Delta m^2\simeq 100\,\mathrm{eV}^2$.

\section{Detection considerations}

We assume the source to be spherical and to be at the center of a spherical detector of radius $R$, the source with its shielding has a radius $r_s$ and thus the event rate is given as
\begin{equation}
    \label{eq:rate}
    n=(R-r_s)\,\sigma\,\rho_t\,D\,,
\end{equation}
where $\sigma$ is the cross section, for the capture reaction it is $5.81\times 10^{-45}\,\mathrm{cm}^2$~\cite{Bahcall:1997eg} and for the elastic scattering reaction it is $5.05 \times 10^{-45}\,\mathrm{cm}^2$~\cite{Tomalak:2019ibg,Hill:2019xqk}.   $D$ is the number of decays in the source $D=\int_T A(t)\,dt$, where $A(t)$ is the source activity as a function of time and $T$ the data taking period.

Gallium by itself does not scintillate, so it needs be incorporated into a scintillating material. We will consider two possibilities here: One is gallium-loaded liquid scintillator; loading of metals into liquid scintillator is a mature technology~\cite{Buck:2016vxe}. However, we found in a detailed analysis (see App.~\ref{sec:consider}) that no combination of plausible liquid scintillator properties and target masses yields a competitive setup.

We therefore will focus on  cerium-doped gadolinium aluminum  gallium garnet (\gaggt), which is a relatively new inorganic scintillator containing 21\% of gallium by weight. {\gaggt} has a high mass density of $6.6\,\mathrm{g}\,\mathrm{cm}^{-3}$, and combined with the  gallium weight fraction this packs a very large number of gallium atoms close to the source. There are  $7.3\times10^{26}$ gallium-71 atoms, assuming the natural isotopic abundance of $^{71}$Ga of 39.89\%, and $2.6\times10^{29}$ electrons per ton of \gaggt. Furthermore, it is a very bright scintillator with light yields ranging from 40,000 -- 60,000~photons per MeV. {\gaggt} has a price of about  \$1,000 per kg and can be manufactured in ton-scale quantities~\cite{quotation}. The attenuation length for scintillation light is reported to be 0.64\,m~\cite{UCHIDA2021164725}. {\gaggt} has a fast scintillation light component of about 100\,ns and a radiation length of 1.6\,cm. It shows good pulse shape discrimination against alpha particles~\cite{KOBAYASHI201291}.

Examples for  light collection systems are given by modern liquid scintillator detectors  like Borexino~\cite{Borexino:2008gab}, JUNO and JUNO-TAO~\cite{JUNO:2015sjr}  with collection efficiencies of approximately 3\%, 10\% and 50\% respectively. JUNO-TAO uses silicon photo multipliers which allow for a complete coverage of the detector surface. These numbers include the photodetector quantum efficiency and light attenuation in the medium. The product of intrinsic light yield and collection efficiency sets the number of detected photo electrons per MeV and this number in turn determines the energy resolution, which is approximately the square root of that number.  We implement a simple optical model in App.~\ref{sec:optics} including attenuation and find for 50\% light collection an effective light yield of approximately 16,000 p.e. per MeV.  The energy resolution is a key figure of merit for the online detection of gallium neutrino capture: the signal of gallium capture is a mono-energetic electron, whereas the leading background arises from elastic neutrino electron scattering, which produces a continuum of electron energies up to $560\,$keV . The rate for elastic scattering is much higher than for gallium capture because the electron density is much higher than the gallium number density. Since both arise from the scattering of neutrinos from the radioactive source, they share their spatial and time distribution.
In a comparison to the solar neutrino rates measured by Borexino~\cite{Borexino:2017rsf,BOREXINO:2020hox,BOREXINO:2020aww} we find that by far the most significant irreducible background to the Ga-capture signal stems from the electron scattering signal from neutrinos from the source.

\section{Germanium tagging}

The very high light yield allows us to consider tagging of the $^{71}$Ge decays which follow with a half-life of 11.4\,d and results in about 88\% of cases in the emission of Auger electrons and/or x-rays with a combined energy of 10.3\,keV. With a detected number of photo electrons of  $p_e=16,000$ per MeV this corresponds around 60 photo electrons, which is well above electronic noise. Borexino has demonstrated a time and space coincidence between the decay of $^{210}$Po with a half-life of 138\,d and the daughter $^{210}$Bi in a liquid environment, which is subject to convection currents~\cite{BOREXINO:2020aww}. Germanium tagging with a half-life of 11.4\,d in a solid should be much easier. The following coincidence criteria are met by Ga-capture events:
\begin{enumerate}
    \item The primary event has an energy of $E_c=510$\,keV, $$\sigma_E = \frac{E_c}{\sqrt{E_c\, p_e}}\simeq 5.6\,\mathrm{keV}\,.$$
    \item The secondary event has an energy of $E_\mathrm{s}=10.3$\,keV $$\sigma_E = \frac{E_\mathrm{s}}{\sqrt{E_\mathrm{s}\,p_e}}\simeq0.8\,\mathrm{keV}\,.$$
     \item The secondary event has to occur at the same position and the resolution is governed by the low-energy secondary event $$\sigma_{r_x} = \frac{r_0}{2}\, \frac{1}{2}\, \frac{1}{\sqrt{E_\mathrm{s}\, p_e}} \simeq 0.8\,\mathrm{cm}\,.$$ 
    \item The secondary event follows with a half-life of 11.4\,d.
\end{enumerate}
Each of these criteria comes with a resolution and selection efficiency. The energy resolution is given by the light yield and respective energies (criteria 1 and 2).  The position resolution (criterion 3) arises from the geometric $1/r^2$ spread. Since for the primary event there is $~50$ times more light, the spatial resolution of the secondary event will determine the coincidence volume.  We expect the leading background to stem from electron scattering from neutrinos from the source. The magnitude of this background can measured in situ by considering energies away from the primary and secondary Ga-capture signal and its energy dependence is fixed by well tested electroweak precision physics. The acceptance of each criterion is determined by how many standard deviations wide each resulting cut is made, with a $1\,\sigma$ range yielding 68\% acceptance, $2\,\sigma$ 95\% asf. In particular, for the spatial criterion~3, there are three powers of this acceptance. We find that choosing the same number of standard deviations for all cuts gives the best sensitivity and the optimum value is around 2.4. Also the coincidence time (criterion~4) can be set in terms of multiples $m_\tau$ of the half-life with a corresponding acceptance of $(1-e^{-m_\tau})$ and we find $m_\tau\sim3$ to be optimal.

A 1.5\,ton detector has a total radius of 40\,cm, including the source with a radius of 15\,cm. We have a volume-average of 16,000 detected photo electrons per MeV with a raw Ga-capture rate per source run of 3.4\,MCi of 210 events and 68,000 electron capture events, which are reduced to about 59,000 events by the coincidence time requirement. The resulting energy spectrum is shown in Fig~\ref{fig:uth}. The primary signal energy criterion leaves 910 background events as primary event candidates. With the spatial resolution of the secondary event of 0.6\,cm applied, this leaves about 10\% of the detector volume as a potential site for the secondary event.  The secondary signal energy criterion then leaves 11 background events which satisfy all 4 criteria. The combined signal acceptance of these criteria and the coincidence time window of 47 days is 88\% for signal events from cuts and 88\% from probability to decay via K-shell capture, whereas the combined background reduction is a factor 7,300 for a final number of 170 signal events over 20 background events. The signal to background ratio is about 1:0.12 compared to the case of no germanium tagging where we found 1:15, or an improvement of  two orders of magnitude. Therefore, the possibility of germanium tagging warrants further study in terms of a more detailed optical model and experimental studies of the low-energy background in {\gaggt} crystals. We note that the required photo detection system would have an active area of about 2.3\,m$^2$, which is about only 1/4th the size of the corresponding system in JUNO-TAO.  These numbers when scaled to 10 source runs yield an overall determination of the Ga-capture rate with an accuracy of 2.7\% and thus, a more than $5\,\sigma$ test of the gallium anomaly.

We further note, that capture to the first excited state of germanium would yield an electron of 370\,keV followed by one gamma photon of 175\,keV 79\,ns later. Given the scintillation time constant of 100\,ns and the 1.6\,cm radiation length~\cite{KOBAYASHI201291} it could be possible to distinguish capture to the first excited state from the one to the ground state and thus, to provide a measurement of this contribution to the cross section; albeit with limited statistics.   
Also, the large sample of neutrino electron scattering events allows for precision tests of the Standard Model as well as search for new physics, like for instance a neutrino magnetic moment~\cite{Coloma:2014hka}.

\section{Backgrounds from the source}

From Eq.~\ref{eq:rate} we can see that the radius of the source $r_s$ has a large impact on the required detector mass to obtain a certain number of events. The BEST source consist of 4.0\,kg of chromium enriched to 97\% of $^{51}$Cr~\cite{Gavrin:2021lam}; given the density of chromium this corresponds to a radius of 5\,cm. This source emits a high rate of 320\,keV gamma photons (branching ratio $\simeq 10\%$). A layered shield of 8\,cm depleted uranium and 2\,cm tungsten, leading to a total source radius of 15\,cm reduces this rate to about 1 event for 10 source runs. Throughout we use the mass energy-absorption coefficients from Ref.~\cite{NIST}. However, the trace impurities in the BEST source chromium become activated during irradiation by neutron capture and result in significant gamma activity at energies between 1--2\,MeV, which is much more difficult to shield. Performing a detailed calculation based on the measured gamma activity and spectrum of the BEST source~\cite{Gavrin:2021lam} indicates that a 40--45\,cm depleted uranium shield is needed to reduce the rate to about 0.1--0.01\,Hz at which point this background is about the same or 1/10th of the intrinsic neutrino electron scattering background. A more detailed simulation would be needed to explore how much of this background ends up in the energy window of the primary and secondary event. An increase of $r_s$ to 45\,cm would require a total detector mass of around 10\,tons to still achieve a 4\% measurement of the Ga-capture rate. There are three mitigation strategies: One is to reduce the impurities of chromium especially those of iron, cobalt, antimony and  titanium from their current sub-ppm level even further. Each factor 10 reduction of those impurities reduces the required shield thickness by about 4.3\,cm and the detector mass by around 1-1.5\,tons. The second mitigation would be to use a $^{37}$Ar source, which would emit only 2\,keV x-rays and Auger electrons and can be made chemically pure without any trace elements as was done for the SAGE experiment~\cite{Abdurashitov:2005tb}. The decay energy is 815\,keV which is very close to the one of $^{51}$Cr of 750\,keV. Production of this source proceeds via the reaction $^{40}$Ca(n,$\alpha$)$^{37}$Ar, which has a reasonable cross section of around 200\,mb but a neutron energy threshold of 2\,MeV,  thus requiring either a fast neutron reactor or an accelerator source of neutrons. The third option is to enrich gallium in gallium-71 up to 80\% from its natural abundance of about 40\%, which largely would compensate the effect the larger source.

\section{Backgrounds in the scintillator}

Borexino has achieved unprecedented levels of radio-purity in its scintillator down to $10^{-17}$g/g~\cite{BOREXINO:2001bob} for uranium-238 and thorium-232 (U/Th), which is the result of a decades-long R\&D effort. For anorganic scintillators much higher numbers are reported in the literature: ranging from as a high as $10^{-10}/10^{-8}$g/g U/Th for  GSO (Gd$_2$SiO$_5$(Ce))~\cite{WANG2002498} down to $10^{-12}$g/g of U/Th for CsI(Tl)~\cite{TEXONO:2000zzq} with similar levels for NaI(Tl)~\cite{PhysRevResearch.2.013223}. The special challenge for {\gaggt} arises from the large content of the rare earth elements gadolonium and cerium. In the context of the R\&D to load Super-K with gadolinium a large-scale process to produce ultra-pure gadolinium sulfate octahydrate has been developed~\cite{superkgd} reaching levels of 2\,ppt of uranium-238 and 0.2\,ppt for thorium-232 for ton quantities of materials.

\begin{figure}[t]
    \centering
    \includegraphics[width=\columnwidth]{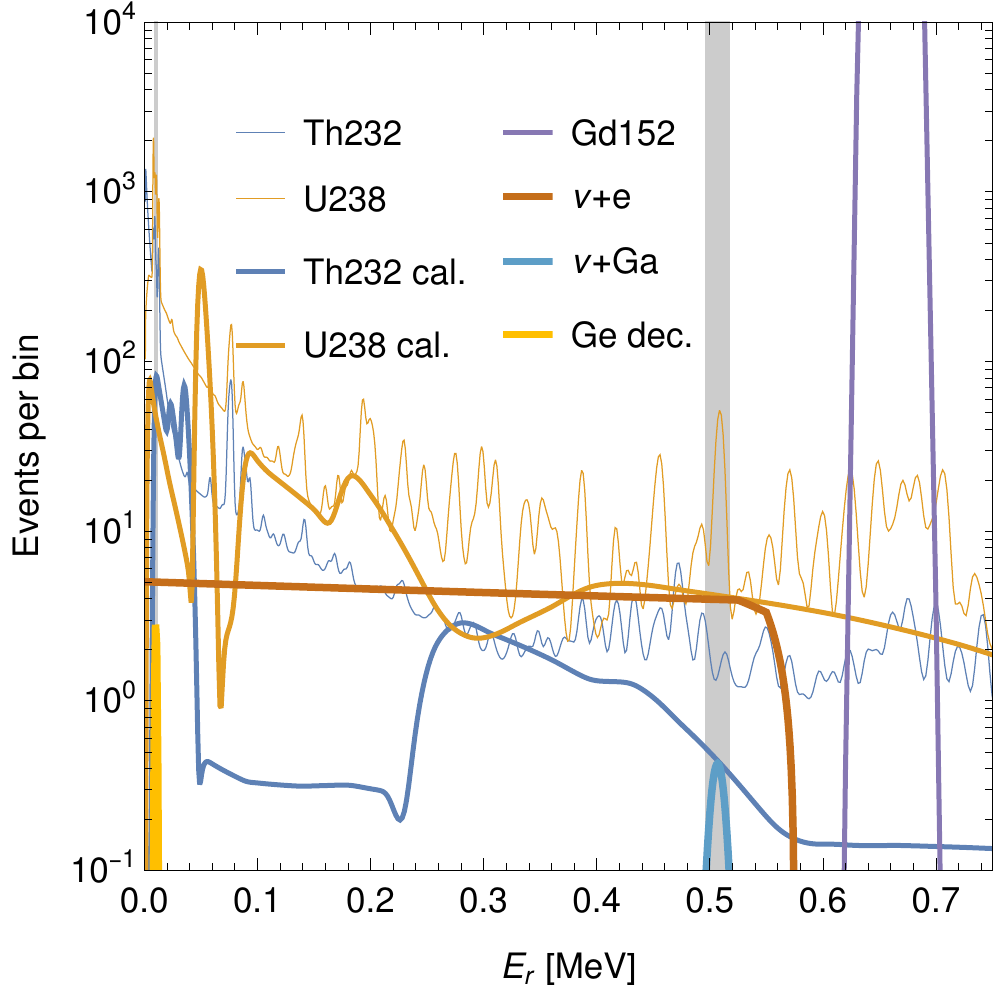}\\
    \includegraphics[width=\columnwidth]{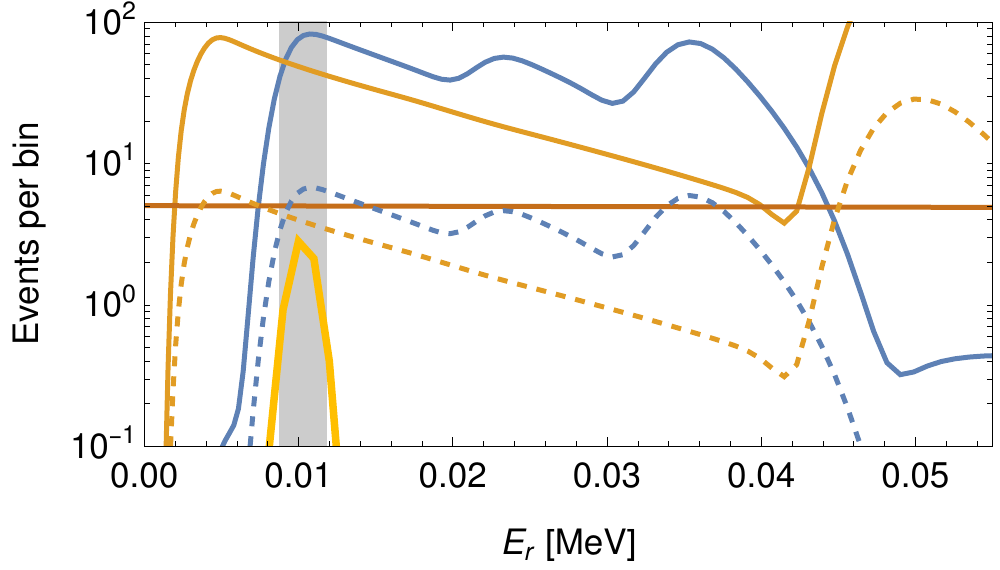}
    
    \caption{Count rate per 1\,keV bin for 80 days of exposure in 1.5\,ton {\gaggt} for a U/Th concentration of $10^{-13}$g/g. We assume secular equilibrium for the decay chains and apply an energy resolution corresponding to an effective light yield of 16,000 p.e./MeV. The grey bands denote the relevant signal energy windows. The lower panel is an enlargement of $0-0.05$\,MeV region, the dashed lines correspond to the remaining background after space and time coincidence cuts with the primary capture event candidate have been applied.}
    \label{fig:uth}
\end{figure}

In Fig~\ref{fig:uth} we show the resulting count rate spectrum for $10^{-13}$g/g of uranium and thorium. We use the decay radiation information, branching fractions and lifetimes from ENSDF~\cite{ensdf} to solve the Bateman equations for the uranium-238 and thorium-232 decay chains and obtain the secular equilibrium abundances to compute the resulting decay radiation. Quenching of $\alpha$-particles in {\gaggt} in the 2-8\,MeV range is found to be around 0.3~\cite{Furuno2021,2018NIMPA}, therefore all $\alpha$-decays are reconstructed above 1\,MeV and thus do not affect this measurement. The thin lines in Fig.\ref{fig:uth} show the result as would be seen in a small crystal or low efficiency detector like a Ge-diode: for each decay only one resulting particle is detected, be it a single gamma or single electron. However, a 1.5\,ton {\gaggt} detector acts a hermetic total electromagnetic calorimeter and thus for any decay all of the available decay energy is deposited with the exception of the energy carried by neutrinos in beta decays. This for instance implies that $\beta^+$-decays do not produce a feature at 511\,keV but at $2\times 511$\,keV plus the kinetic energy of the positron. Also Auger electrons and atomic x-rays do not show up at 10s of keV but around the Q-value of the decay. To illustrate this effect we compute the total electromagnetic energy deposition where we account for the neutrino energy using the framework of Ref.~\cite{Huber:2011wv}. The result is shown as thick lines in Fig.~\ref{fig:uth} and we observe a reduction of 12.3 in the low-energy signal window and 5.4 in the high-energy one relative to the low-efficency detector case. In particular the thallium-208 gamma lines at 510\,keV and 583\,keV are removed since they now get combined with all gammas from this decay which add up to more than 4\,MeV. 
Using the result for a calorimetric detector we now can repeat the previous germanium tagging analysis including the U/Th background assuming equal concentrations of both and find that a 4\% measurement of the gallium capture signal is feasible with a 1.5\,ton detector for a U/Th concentration of $1.3\times 10^{-12}$g/g each. This is significantly lower than the concentration found in a commercial GSO detector but falls is within the range obtained with a directed effort at improved radiopurity for CsI(Tl)~\cite{TEXONO:2000zzq} and NaI(Tl) scintillators~\cite{PhysRevResearch.2.013223}. In Fig.\ref{fig:uth} we show the background spectrum for $10^{-13}$g/g of U/Th. Combined with the Ge-tagging coincidence cuts, the effective background level in the signal regions does increase only moderately; we still would obtain a measurement at 3.6\% precision.

Another intrinsic background arises from gadolinium-152 which has natural abundance of 0.2\% and a half-life of $1.08\times10^{14}$\,y and decays with the emission of a 2.2\,MeV $\alpha$-particle. With a 0.3 quenching factor this becomes 660\,keV reconstructed energy and is thus far enough from the signal region at 500\,keV so that it does not affect the background estimate as can be seen from Fig.~\ref{fig:uth}.

There are two stable gadolinium isotopes with mass numbers 155 and 157, with sizable abundances above 10\% with large neutron capture cross sections of $6\times10^4$\,barn and $2.5\times10^5$\,barn, respectively, which both however will capture to other stable isotopes. In both cases multi-MeV gamma cascades result which are too high in energy to matter here.

Cosmogenic backgrounds of concern are for instance germanium-68 (270\,d half-life) and tritium (12\,y half-life), both low energy decays with 10\,keV and 5.6\,keV mean energy. The half-life of tritium is such that decay will not ameliorate the problem and the experiment can only tolerate about 1,000 atoms of tritium per kilogram. Typical cosmogenic production rates at sea level are of the order 100 atoms per day and kilogram~\cite{Amare:2017roa,Saldanha:2020ubf} and this clearly indicates the need to grow the crystals underground.

These initial theoretical estimates are no substitute for an experimental program to study the actual background levels and to initiate R\&D towards radio-pure {\gaggt}. These results do however indicate that the challenge may not be insurmountable, in particular Borexino-level radio-purity is not required.

\section{Conclusion}

We argue that a ton-scale scintillation experiment using {\gaggt} combined with several runs of mega-Curie strength $^{51}$Cr sources can provide a direct test of the gallium anomaly at the $5\,\sigma$ level by individually addressing each potential root cause. In particular, the comparison of the neutrino electron scattering rate to the Ga-capture rate avoids the potential systematics issue which may have been encountered by previous measurements and would give insights into the underlying cause of the gallium anomaly. We believe these results warrant a more in-depth investigation of the feasibility of such an experiment.

\section*{Acknowledgements}

 I would like to thank M.~Yeh for useful conversations regarding the loading of gallium into liquid scintillator, J.~Berryman for comments on the manuscript and J.M.~Link for in-depth feedback. I would like to thank M.~Cribier for bringing gadolinium-152 to my attention and an anonymous referee for sharpening the discussion of backgrounds. The author would also like to thank the Instituto de Fisica Teorica (IFT UAM-CSIC) in Madrid for support via the Centro de Excelencia Severo Ochoa Program under Grant CEX2020-001007-S, during the Extended Workshop “Neutrino Theories”, where this work developed. The work was supported by the
U.S. Department of Energy Office of Science under award number
DE-SC00018327.

\begin{appendix}

\section{Experimental considerations}
\label{sec:consider}

We did study gallium loaded liquid scintillator in detail. The advantages of Ga-loaded scintillator are that existing detectors can be repurposed and the overall cost is relatively low. On the other hand, the  gallium fraction is limited, and increasing the gallium fraction beyond a certain point would lead to a strong drop in light yield and transparency. Also, the mass density of liquid scintillator is relatively low, which when combined with the limited gallium fraction leads to a relatively low density of gallium atoms. The light yields for Ga-loaded liquid scintillator may range from 3,000 -- 8,000 photons/MeV with attenuation length anywhere from 2--8\,m.

The closest example of an existing highly metal loaded scintillator is the one developed for the LENS experiment, which achieved an indium loading of 8 to 10\% with a light yield of 5,500 photons per MeV, and a long-term stable attenuation length of 8\,m~\cite{raghavan2008lens}. It is unclear if this specific recipe allows one to increase indium loading to 20\% while maintaining the light yield and attenuation length. Modern, water-based liquid scintillator formulations could allow for high loading fractions of 20\% with light yields in the range of 3,000-4,200 photons per MeV and attenuation length of 1-2\,m. Then there is also the option to use trimethylgallium (TMG), which is pyrophoric, corrosive and rather expensive.  TMG can be mixed directly with scintillator, and it appears plausible that light yields around 8,000 photons per MeV and an attenuation length of 10\,m or more can be  achieved. The resulting number of detected photo electrons per MeV makes gallium-tagging impossible.

We perform a log-likelihood analysis with data binned into 1\,keV bins in the energy range from 0.2--0.8\,MeV and consider the rates from source induced electron scattering and gallium capture events as well as the background from $^{210}$Po. In this analysis the distinction of Ga-capture events from electron scattering events stems purely from the energy spectrum and no Ga-tagging is considered.

We include an optical model for the attenuation by averaging the attenuation across the whole volume and we do account for shadowing of the source which is assumed to be a sphere of 15\,cm radius. For the 
{\gaggt} case we assume that the surface of the source is also photosensitive which helps to mitigate the otherwise large shadowing effect, see App.~\ref{sec:optics}. 

We next study the required detector mass, which is needed to achieve 4\% precision on the Ga-capture rate or a $5\,\sigma$ test of the gallium anomaly. We evaluate the mass as a function of the effective light yield which is the product of the intrinsic scintillator light yield and the detection efficiency, which in turn is the product of quantum efficiency and photo cathode coverage, and light survival probability, set by the attenuation length of the scintillator.  For no plausible combination of liquid scintillator properties  could we find a configuration which can provide 4\% precision with 25,000\,tons or less of detector mass, even with a light collection efficiency of 0.5. For {\gaggt} the specific scintillator properties matter much less and we find detector masses in the range of 25 to 40\,ton corresponding to a light detection system only 4 times larger than, but otherwise with the same performance, the one of JUNO-TAO. The main challenge here would be manufacture of such a large amount of \gaggt. The resulting Ga-capture signal event rate is around 10,000 events compared to an overall background from electron scattering of 3.2 million events. The energy resolution effectively reduces the electron scattering background to around 150,000 events in the Ga-capture signal region corresponding to a signal to background ratio of 1:15. The required detector mass is relatively large given the cost of the scintillator.

We therefore conclude, that liquid scintillator is not a viable option and that Ga-tagging is the key to obtain the best possible performance.

\section{Optical model}
\label{sec:optics}

\begin{figure}
    \centering
    \includegraphics[width=\columnwidth]{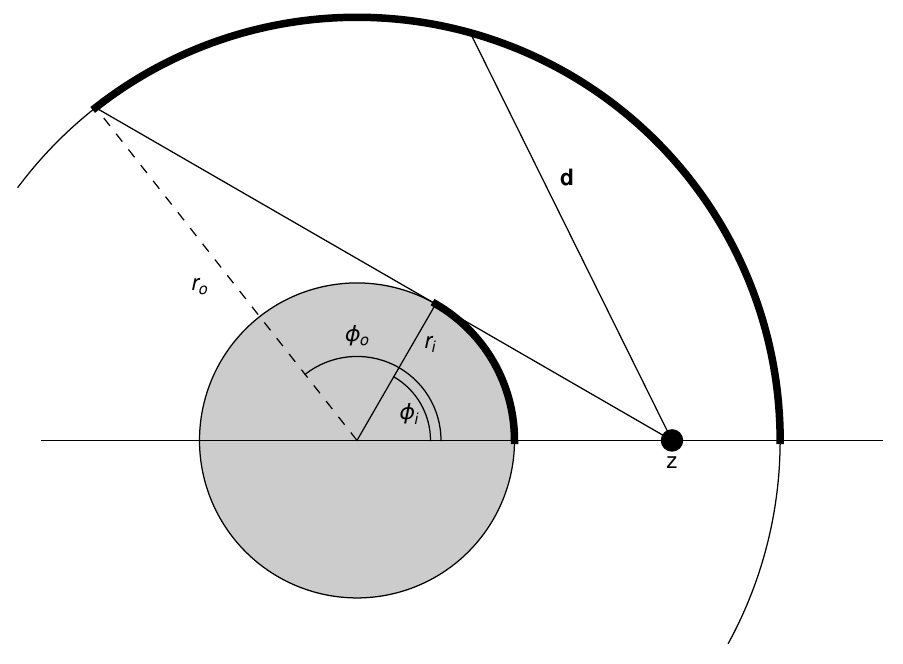}
    \caption{Schematic of a spherical detector with both inner and outer surface being instrumented. The center gray circle depicts the source and its shield.}
    \label{fig:opticalmodel}
\end{figure}

In order to include the effect of light attenuation in the scintillator we employ the following simple geometrical model, as shown in Fig.~\ref{fig:opticalmodel}: two concentric spheres with radii $r_i$ and $r_o$ respectively, representing the source with its shield and the outer surface of the detector. We further assume that both the surface of the source assembly and the outer surface are fully covered with photosensors, e.g. SiPMs as in JUNO TAO. For any point in the interior we can define a coordinate system where the z-axis goes from the origin through this point. We further can construct the intervals $[0,\phi_i]$ where the light will be hitting the inner surface and $[0,\phi_o]$ where light will be hitting the outer surface.
We the can compute the resulting fraction of light collected by each surface using the following function,
\begin{eqnarray}
    f(r,\phi_c,z)&=&\nonumber\\  \int_0^{\phi_c}\,\sin \phi\, \mathrm{d}\phi\,\mathrm{d} \theta&&\,\frac{r}{4\pi\,|\mathbf{d}(r,\phi,\theta)|}\exp\left(-\frac{|\mathbf{d}(r,\phi,\theta)|}{\lambda}\right) \,,\nonumber\\
    \mathbf{d(}r,\phi,\theta)&=&\left(\begin{array}{c}0\\0\\ z\end{array}\right)-r\left(\begin{array}{c}\sin\phi\,\sin\theta\\\sin\phi\,\cos\theta\\ \cos\phi\end{array}\right)\,,
\end{eqnarray}
which does account for attenuation. Next we average this result over the whole volume recognizing that the signal event rate is constant in $z$ (the radial direction), yielding the following final result:
\begin{equation}
    F=\frac{1}{r_o-r_i}\int_{r_i}^{r_o} \,\mathrm{d}z\,f(r_i,\phi_i,z)+f(r_o,\phi_o,z)\,.
\end{equation}
Note, that backgrounds scale like $z^2$ and hence will be on average closer to the outer surface and have somewhat larger values for $F$, which we will neglect here. In an actual experiment the position of the event is known and a fully position dependent resolution function can be used event-by-event.

\begin{figure}
    \centering
    \includegraphics[width=\columnwidth]{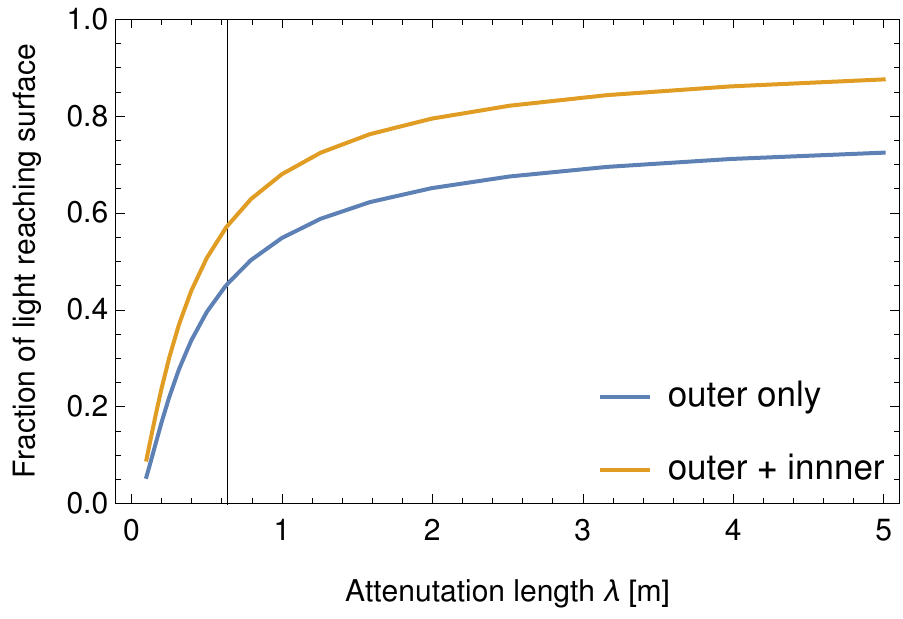}
    \caption{The fraction $F$ of light reaching the surface of the detector as a function of the attenuation length with $r_i=0.15\,m$ and $r_0=0.403$\,m corresponding to 1.5 ton target mass. The vertical line denotes the default attenuation length $\lambda=0.64$\,m. }
    \label{fig:optics}
\end{figure}

For our simplified model the following result, as shown in Fig.~\ref{fig:optics} is found for using the nominal values of $r_i=0.15$\,m and $r_o=0.403$\,m. Also shown is the result for light collection only along the outer surface (blue line) versus both surfaces instrumented. For the results in the main body of the paper we use an attenuation length of $\lambda=0.64$\,m~\cite{KOBAYASHI201291} yielding a light collection fraction of 57.4\% using both surfaces.

\end{appendix}

%\clearpage
\bibliography{apssamp}% Produces the bibliography via BibTeX.

%apsrev4-2.bst 2019-01-14 (MD) hand-edited version of apsrev4-1.bst
%Control: key (0)
%Control: author (8) initials jnrlst
%Control: editor formatted (1) identically to author
%Control: production of article title (0) allowed
%Control: page (0) single
%Control: year (1) truncated
%Control: production of eprint (0) enabled
\providecommand{\noopsort}[1]{}\providecommand{\singleletter}[1]{#1}%
\begin{thebibliography}{40}%
\makeatletter
\providecommand \@ifxundefined [1]{%
 \@ifx{#1\undefined}
}%
\providecommand \@ifnum [1]{%
 \ifnum #1\expandafter \@firstoftwo
 \else \expandafter \@secondoftwo
 \fi
}%
\providecommand \@ifx [1]{%
 \ifx #1\expandafter \@firstoftwo
 \else \expandafter \@secondoftwo
 \fi
}%
\providecommand \natexlab [1]{#1}%
\providecommand \enquote  [1]{``#1''}%
\providecommand \bibnamefont  [1]{#1}%
\providecommand \bibfnamefont [1]{#1}%
\providecommand \citenamefont [1]{#1}%
\providecommand \href@noop [0]{\@secondoftwo}%
\providecommand \href [0]{\begingroup \@sanitize@url \@href}%
\providecommand \@href[1]{\@@startlink{#1}\@@href}%
\providecommand \@@href[1]{\endgroup#1\@@endlink}%
\providecommand \@sanitize@url [0]{\catcode `\\12\catcode `\$12\catcode
  `\&12\catcode `\#12\catcode `\^12\catcode `\_12\catcode `\%12\relax}%
\providecommand \@@startlink[1]{}%
\providecommand \@@endlink[0]{}%
\providecommand \url  [0]{\begingroup\@sanitize@url \@url }%
\providecommand \@url [1]{\endgroup\@href {#1}{\urlprefix }}%
\providecommand \urlprefix  [0]{URL }%
\providecommand \Eprint [0]{\href }%
\providecommand \doibase [0]{https://doi.org/}%
\providecommand \selectlanguage [0]{\@gobble}%
\providecommand \bibinfo  [0]{\@secondoftwo}%
\providecommand \bibfield  [0]{\@secondoftwo}%
\providecommand \translation [1]{[#1]}%
\providecommand \BibitemOpen [0]{}%
\providecommand \bibitemStop [0]{}%
\providecommand \bibitemNoStop [0]{.\EOS\space}%
\providecommand \EOS [0]{\spacefactor3000\relax}%
\providecommand \BibitemShut  [1]{\csname bibitem#1\endcsname}%
\let\auto@bib@innerbib\@empty
%</preamble>
\bibitem [{\citenamefont {Abdurashitov}\ \emph {et~al.}(1999)\citenamefont
  {Abdurashitov} \emph {et~al.}}]{Abdurashitov:1998ne}%
  \BibitemOpen
  \bibfield  {author} {\bibinfo {author} {\bibfnamefont {J.~N.}\ \bibnamefont
  {Abdurashitov}} \emph {et~al.} (\bibinfo {collaboration} {SAGE}),\ }\bibfield
   {title} {\bibinfo {title} {{Measurement of the response of the
  Russian-American gallium experiment to neutrinos from a Cr-51 source}},\
  }\href {https://doi.org/10.1103/PhysRevC.59.2246} {\bibfield  {journal}
  {\bibinfo  {journal} {Phys. Rev.}\ }\textbf {\bibinfo {volume} {C59}},\
  \bibinfo {pages} {2246} (\bibinfo {year} {1999})},\ \Eprint
  {https://arxiv.org/abs/hep-ph/9803418} {arXiv:hep-ph/9803418 [hep-ph]}
  \BibitemShut {NoStop}%
%%CITATION = HEP-PH/9803418;%%
\bibitem [{\citenamefont {Hampel}\ \emph {et~al.}(1998)\citenamefont {Hampel}
  \emph {et~al.}}]{Hampel:1997fc}%
  \BibitemOpen
  \bibfield  {author} {\bibinfo {author} {\bibfnamefont {W.}~\bibnamefont
  {Hampel}} \emph {et~al.} (\bibinfo {collaboration} {GALLEX}),\ }\bibfield
  {title} {\bibinfo {title} {{Final results of the Cr-51 neutrino source
  experiments in GALLEX}},\ }\href
  {https://doi.org/10.1016/S0370-2693(97)01562-1} {\bibfield  {journal}
  {\bibinfo  {journal} {Phys. Lett.}\ }\textbf {\bibinfo {volume} {B420}},\
  \bibinfo {pages} {114} (\bibinfo {year} {1998})}\BibitemShut {NoStop}%
%%CITATION = PHLTA,B420,114;%%
\bibitem [{\citenamefont {Kaether}\ \emph {et~al.}(2010)\citenamefont
  {Kaether}, \citenamefont {Hampel}, \citenamefont {Heusser}, \citenamefont
  {Kiko},\ and\ \citenamefont {Kirsten}}]{Kaether:2010ag}%
  \BibitemOpen
  \bibfield  {author} {\bibinfo {author} {\bibfnamefont {F.}~\bibnamefont
  {Kaether}}, \bibinfo {author} {\bibfnamefont {W.}~\bibnamefont {Hampel}},
  \bibinfo {author} {\bibfnamefont {G.}~\bibnamefont {Heusser}}, \bibinfo
  {author} {\bibfnamefont {J.}~\bibnamefont {Kiko}},\ and\ \bibinfo {author}
  {\bibfnamefont {T.}~\bibnamefont {Kirsten}},\ }\bibfield  {title} {\bibinfo
  {title} {{Reanalysis of the GALLEX solar neutrino flux and source
  experiments}},\ }\href {https://doi.org/10.1016/j.physletb.2010.01.030}
  {\bibfield  {journal} {\bibinfo  {journal} {Phys. Lett.}\ }\textbf {\bibinfo
  {volume} {B685}},\ \bibinfo {pages} {47} (\bibinfo {year} {2010})},\ \Eprint
  {https://arxiv.org/abs/1001.2731} {arXiv:1001.2731 [hep-ex]} \BibitemShut
  {NoStop}%
%%CITATION = ARXIV:1001.2731;%%
\bibitem [{\citenamefont {Barinov}\ \emph {et~al.}(2021)\citenamefont {Barinov}
  \emph {et~al.}}]{Barinov:2021asz}%
  \BibitemOpen
  \bibfield  {author} {\bibinfo {author} {\bibfnamefont {V.~V.}\ \bibnamefont
  {Barinov}} \emph {et~al.},\ }\bibfield  {title} {\bibinfo {title} {{Results
  from the Baksan Experiment on Sterile Transitions (BEST)}},\ }\href@noop {}
  {\  (\bibinfo {year} {2021})},\ \Eprint {https://arxiv.org/abs/2109.11482}
  {arXiv:2109.11482 [nucl-ex]} \BibitemShut {NoStop}%
\bibitem [{\citenamefont {Abdurashitov}\ \emph {et~al.}(2006)\citenamefont
  {Abdurashitov} \emph {et~al.}}]{Abdurashitov:2005tb}%
  \BibitemOpen
  \bibfield  {author} {\bibinfo {author} {\bibfnamefont {J.~N.}\ \bibnamefont
  {Abdurashitov}} \emph {et~al.},\ }\bibfield  {title} {\bibinfo {title}
  {{Measurement of the response of a Ga solar neutrino experiment to neutrinos
  from an Ar-37 source}},\ }\href {https://doi.org/10.1103/PhysRevC.73.045805}
  {\bibfield  {journal} {\bibinfo  {journal} {Phys. Rev.}\ }\textbf {\bibinfo
  {volume} {C73}},\ \bibinfo {pages} {045805} (\bibinfo {year} {2006})},\
  \Eprint {https://arxiv.org/abs/nucl-ex/0512041} {arXiv:nucl-ex/0512041
  [nucl-ex]} \BibitemShut {NoStop}%
%%CITATION = NUCL-EX/0512041;%%
\bibitem [{\citenamefont {Berryman}\ \emph {et~al.}(2022)\citenamefont
  {Berryman}, \citenamefont {Coloma}, \citenamefont {Huber}, \citenamefont
  {Schwetz},\ and\ \citenamefont {Zhou}}]{Berryman:2021yan}%
  \BibitemOpen
  \bibfield  {author} {\bibinfo {author} {\bibfnamefont {J.~M.}\ \bibnamefont
  {Berryman}}, \bibinfo {author} {\bibfnamefont {P.}~\bibnamefont {Coloma}},
  \bibinfo {author} {\bibfnamefont {P.}~\bibnamefont {Huber}}, \bibinfo
  {author} {\bibfnamefont {T.}~\bibnamefont {Schwetz}},\ and\ \bibinfo {author}
  {\bibfnamefont {A.}~\bibnamefont {Zhou}},\ }\bibfield  {title} {\bibinfo
  {title} {{Statistical significance of the sterile-neutrino hypothesis in the
  context of reactor and gallium data}},\ }\href
  {https://doi.org/10.1007/JHEP02(2022)055} {\bibfield  {journal} {\bibinfo
  {journal} {JHEP}\ }\textbf {\bibinfo {volume} {02}},\ \bibinfo {pages}
  {055}},\ \Eprint {https://arxiv.org/abs/2111.12530} {arXiv:2111.12530
  [hep-ph]} \BibitemShut {NoStop}%
\bibitem [{\citenamefont {Haxton}\ \emph {et~al.}(2023)\citenamefont {Haxton},
  \citenamefont {Rule}, \citenamefont {Elliott}, \citenamefont {Gavrin},\ and\
  \citenamefont {Ibragimova}}]{Haxton:2023ppq}%
  \BibitemOpen
  \bibfield  {author} {\bibinfo {author} {\bibfnamefont {W.~C.}\ \bibnamefont
  {Haxton}}, \bibinfo {author} {\bibfnamefont {E.~J.}\ \bibnamefont {Rule}},
  \bibinfo {author} {\bibfnamefont {S.~R.}\ \bibnamefont {Elliott}}, \bibinfo
  {author} {\bibfnamefont {V.~N.}\ \bibnamefont {Gavrin}},\ and\ \bibinfo
  {author} {\bibfnamefont {T.~V.}\ \bibnamefont {Ibragimova}},\ }\bibfield
  {title} {\bibinfo {title} {{The Gallium Neutrino Absorption Cross Section and
  its Uncertainty}},\ }\href@noop {} {\  (\bibinfo {year} {2023})},\ \Eprint
  {https://arxiv.org/abs/2303.13623} {arXiv:2303.13623 [nucl-th]} \BibitemShut
  {NoStop}%
\bibitem [{\citenamefont {Brdar}\ \emph {et~al.}(2023)\citenamefont {Brdar},
  \citenamefont {Gehrlein},\ and\ \citenamefont {Kopp}}]{Brdar:2023cms}%
  \BibitemOpen
  \bibfield  {author} {\bibinfo {author} {\bibfnamefont {V.}~\bibnamefont
  {Brdar}}, \bibinfo {author} {\bibfnamefont {J.}~\bibnamefont {Gehrlein}},\
  and\ \bibinfo {author} {\bibfnamefont {J.}~\bibnamefont {Kopp}},\ }\bibfield
  {title} {\bibinfo {title} {{Towards Resolving the Gallium Anomaly}},\
  }\href@noop {} {\  (\bibinfo {year} {2023})},\ \Eprint
  {https://arxiv.org/abs/2303.05528} {arXiv:2303.05528 [hep-ph]} \BibitemShut
  {NoStop}%
\bibitem [{\citenamefont {Raghavan}(1976)}]{Raghavan:1976yc}%
  \BibitemOpen
  \bibfield  {author} {\bibinfo {author} {\bibfnamefont {R.~S.}\ \bibnamefont
  {Raghavan}},\ }\bibfield  {title} {\bibinfo {title} {{Inverse beta decay of
  115-In to 115-Sn*: a new possibility for detecting solar neutrinos from the
  proton-proton reaction}},\ }\href
  {https://doi.org/10.1103/PhysRevLett.37.259} {\bibfield  {journal} {\bibinfo
  {journal} {Phys. Rev. Lett.}\ }\textbf {\bibinfo {volume} {37}},\ \bibinfo
  {pages} {259} (\bibinfo {year} {1976})}\BibitemShut {NoStop}%
\bibitem [{\citenamefont {Grieb}\ \emph {et~al.}(2007)\citenamefont {Grieb},
  \citenamefont {Link},\ and\ \citenamefont {Raghavan}}]{Grieb:2006mp}%
  \BibitemOpen
  \bibfield  {author} {\bibinfo {author} {\bibfnamefont {C.}~\bibnamefont
  {Grieb}}, \bibinfo {author} {\bibfnamefont {J.}~\bibnamefont {Link}},\ and\
  \bibinfo {author} {\bibfnamefont {R.~S.}\ \bibnamefont {Raghavan}},\
  }\bibfield  {title} {\bibinfo {title} {{Probing active to sterile neutrino
  oscillations in the LENS detector}},\ }\href
  {https://doi.org/10.1103/PhysRevD.75.093006} {\bibfield  {journal} {\bibinfo
  {journal} {Phys. Rev. D}\ }\textbf {\bibinfo {volume} {75}},\ \bibinfo
  {pages} {093006} (\bibinfo {year} {2007})},\ \Eprint
  {https://arxiv.org/abs/hep-ph/0611178} {arXiv:hep-ph/0611178} \BibitemShut
  {NoStop}%
\bibitem [{\citenamefont {Bellini}\ \emph {et~al.}(2013)\citenamefont {Bellini}
  \emph {et~al.}}]{Borexino:2013xxa}%
  \BibitemOpen
  \bibfield  {author} {\bibinfo {author} {\bibfnamefont {G.}~\bibnamefont
  {Bellini}} \emph {et~al.} (\bibinfo {collaboration} {Borexino}),\ }\bibfield
  {title} {\bibinfo {title} {{SOX: Short distance neutrino Oscillations with
  BoreXino}},\ }\href {https://doi.org/10.1007/JHEP08(2013)038} {\bibfield
  {journal} {\bibinfo  {journal} {JHEP}\ }\textbf {\bibinfo {volume} {08}},\
  \bibinfo {pages} {038}},\ \Eprint {https://arxiv.org/abs/1304.7721}
  {arXiv:1304.7721 [physics.ins-det]} \BibitemShut {NoStop}%
\bibitem [{\citenamefont {Tomalak}\ and\ \citenamefont
  {Hill}(2020)}]{Tomalak:2019ibg}%
  \BibitemOpen
  \bibfield  {author} {\bibinfo {author} {\bibfnamefont {O.}~\bibnamefont
  {Tomalak}}\ and\ \bibinfo {author} {\bibfnamefont {R.~J.}\ \bibnamefont
  {Hill}},\ }\bibfield  {title} {\bibinfo {title} {{Theory of elastic
  neutrino-electron scattering}},\ }\href
  {https://doi.org/10.1103/PhysRevD.101.033006} {\bibfield  {journal} {\bibinfo
   {journal} {Phys. Rev. D}\ }\textbf {\bibinfo {volume} {101}},\ \bibinfo
  {pages} {033006} (\bibinfo {year} {2020})},\ \Eprint
  {https://arxiv.org/abs/1907.03379} {arXiv:1907.03379 [hep-ph]} \BibitemShut
  {NoStop}%
\bibitem [{\citenamefont {Hill}\ and\ \citenamefont
  {Tomalak}(2020)}]{Hill:2019xqk}%
  \BibitemOpen
  \bibfield  {author} {\bibinfo {author} {\bibfnamefont {R.~J.}\ \bibnamefont
  {Hill}}\ and\ \bibinfo {author} {\bibfnamefont {O.}~\bibnamefont {Tomalak}},\
  }\bibfield  {title} {\bibinfo {title} {{On the effective theory of
  neutrino-electron and neutrino-quark interactions}},\ }\href
  {https://doi.org/10.1016/j.physletb.2020.135466} {\bibfield  {journal}
  {\bibinfo  {journal} {Phys. Lett. B}\ }\textbf {\bibinfo {volume} {805}},\
  \bibinfo {pages} {135466} (\bibinfo {year} {2020})},\ \Eprint
  {https://arxiv.org/abs/1911.01493} {arXiv:1911.01493 [hep-ph]} \BibitemShut
  {NoStop}%
\bibitem [{\citenamefont {Giunti}\ and\ \citenamefont
  {Laveder}(2011)}]{Giunti:2010zu}%
  \BibitemOpen
  \bibfield  {author} {\bibinfo {author} {\bibfnamefont {C.}~\bibnamefont
  {Giunti}}\ and\ \bibinfo {author} {\bibfnamefont {M.}~\bibnamefont
  {Laveder}},\ }\bibfield  {title} {\bibinfo {title} {{Statistical Significance
  of the Gallium Anomaly}},\ }\href
  {https://doi.org/10.1103/PhysRevC.83.065504} {\bibfield  {journal} {\bibinfo
  {journal} {Phys. Rev. C}\ }\textbf {\bibinfo {volume} {83}},\ \bibinfo
  {pages} {065504} (\bibinfo {year} {2011})},\ \Eprint
  {https://arxiv.org/abs/1006.3244} {arXiv:1006.3244 [hep-ph]} \BibitemShut
  {NoStop}%
\bibitem [{\citenamefont {Giunti}\ \emph {et~al.}(2022)\citenamefont {Giunti},
  \citenamefont {Li}, \citenamefont {Ternes}, \citenamefont {Tyagi},\ and\
  \citenamefont {Xin}}]{Giunti:2022btk}%
  \BibitemOpen
  \bibfield  {author} {\bibinfo {author} {\bibfnamefont {C.}~\bibnamefont
  {Giunti}}, \bibinfo {author} {\bibfnamefont {Y.~F.}\ \bibnamefont {Li}},
  \bibinfo {author} {\bibfnamefont {C.~A.}\ \bibnamefont {Ternes}}, \bibinfo
  {author} {\bibfnamefont {O.}~\bibnamefont {Tyagi}},\ and\ \bibinfo {author}
  {\bibfnamefont {Z.}~\bibnamefont {Xin}},\ }\bibfield  {title} {\bibinfo
  {title} {{Gallium Anomaly: Critical View from the Global Picture of $\nu_{e}$
  and $\bar\nu_{e}$ Disappearance}},\ }\href@noop {} {\  (\bibinfo {year}
  {2022})},\ \Eprint {https://arxiv.org/abs/2209.00916} {arXiv:2209.00916
  [hep-ph]} \BibitemShut {NoStop}%
\bibitem [{\citenamefont {Bahcall}(1997)}]{Bahcall:1997eg}%
  \BibitemOpen
  \bibfield  {author} {\bibinfo {author} {\bibfnamefont {J.~N.}\ \bibnamefont
  {Bahcall}},\ }\bibfield  {title} {\bibinfo {title} {{Gallium solar neutrino
  experiments: Absorption cross-sections, neutrino spectra, and predicted event
  rates}},\ }\href {https://doi.org/10.1103/PhysRevC.56.3391} {\bibfield
  {journal} {\bibinfo  {journal} {Phys. Rev. C}\ }\textbf {\bibinfo {volume}
  {56}},\ \bibinfo {pages} {3391} (\bibinfo {year} {1997})},\ \Eprint
  {https://arxiv.org/abs/hep-ph/9710491} {arXiv:hep-ph/9710491} \BibitemShut
  {NoStop}%
\bibitem [{\citenamefont {Buck}\ and\ \citenamefont
  {Yeh}(2016)}]{Buck:2016vxe}%
  \BibitemOpen
  \bibfield  {author} {\bibinfo {author} {\bibfnamefont {C.}~\bibnamefont
  {Buck}}\ and\ \bibinfo {author} {\bibfnamefont {M.}~\bibnamefont {Yeh}},\
  }\bibfield  {title} {\bibinfo {title} {{Metal-loaded organic scintillators
  for neutrino physics}},\ }\href
  {https://doi.org/10.1088/0954-3899/43/9/093001} {\bibfield  {journal}
  {\bibinfo  {journal} {J. Phys. G}\ }\textbf {\bibinfo {volume} {43}},\
  \bibinfo {pages} {093001} (\bibinfo {year} {2016})},\ \Eprint
  {https://arxiv.org/abs/1608.04897} {arXiv:1608.04897 [physics.ins-det]}
  \BibitemShut {NoStop}%
\bibitem [{quo()}]{quotation}%
  \BibitemOpen
  \bibinfo {note} {Vendor quote from Hangzhou Shalom Electro-optics Technology
  Co., Ltd.}\BibitemShut {Stop}%
\bibitem [{\citenamefont {Uchida}\ \emph {et~al.}(2021)\citenamefont {Uchida},
  \citenamefont {Takahashi}, \citenamefont {Ohno}, \citenamefont {Mizuno},
  \citenamefont {Fukazawa}, \citenamefont {Yoshino}, \citenamefont {Kamada},
  \citenamefont {Yokota},\ and\ \citenamefont {Yoshikawa}}]{UCHIDA2021164725}%
  \BibitemOpen
  \bibfield  {author} {\bibinfo {author} {\bibfnamefont {N.}~\bibnamefont
  {Uchida}}, \bibinfo {author} {\bibfnamefont {H.}~\bibnamefont {Takahashi}},
  \bibinfo {author} {\bibfnamefont {M.}~\bibnamefont {Ohno}}, \bibinfo {author}
  {\bibfnamefont {T.}~\bibnamefont {Mizuno}}, \bibinfo {author} {\bibfnamefont
  {Y.}~\bibnamefont {Fukazawa}}, \bibinfo {author} {\bibfnamefont
  {M.}~\bibnamefont {Yoshino}}, \bibinfo {author} {\bibfnamefont
  {K.}~\bibnamefont {Kamada}}, \bibinfo {author} {\bibfnamefont
  {Y.}~\bibnamefont {Yokota}},\ and\ \bibinfo {author} {\bibfnamefont
  {A.}~\bibnamefont {Yoshikawa}},\ }\bibfield  {title} {\bibinfo {title}
  {Attenuation characteristics of a \protect{Ce:Gd$_3$Al$_2$Ga$_3$O$_{12}$}
  scintillator},\ }\href
  {https://doi.org/https://doi.org/10.1016/j.nima.2020.164725} {\bibfield
  {journal} {\bibinfo  {journal} {Nuclear Instruments and Methods in Physics
  Research Section A: Accelerators, Spectrometers, Detectors and Associated
  Equipment}\ }\textbf {\bibinfo {volume} {986}},\ \bibinfo {pages} {164725}
  (\bibinfo {year} {2021})}\BibitemShut {NoStop}%
\bibitem [{KOB(2012)}]{KOBAYASHI201291}%
  \BibitemOpen
  \bibfield  {title} {\bibinfo {title} {Significantly different pulse shapes
  for $\gamma$- and $\alpha$-rays in
  \protect{Gd$_3$Al$_2$Ga$_3$O$_{12}$:Ce$^{3+}$} scintillating crystals},\
  }\href@noop {} {\bibfield  {journal} {\bibinfo  {journal} {Nuclear
  Instruments and Methods in Physics Research Section A: Accelerators,
  Spectrometers, Detectors and Associated Equipment}\ }\textbf {\bibinfo
  {volume} {694}},\ \bibinfo {pages} {91} (\bibinfo {year} {2012})}\BibitemShut
  {NoStop}%
\bibitem [{\citenamefont {Alimonti}\ \emph {et~al.}(2009)\citenamefont
  {Alimonti} \emph {et~al.}}]{Borexino:2008gab}%
  \BibitemOpen
  \bibfield  {author} {\bibinfo {author} {\bibfnamefont {G.}~\bibnamefont
  {Alimonti}} \emph {et~al.} (\bibinfo {collaboration} {Borexino}),\ }\bibfield
   {title} {\bibinfo {title} {{The Borexino detector at the Laboratori
  Nazionali del Gran Sasso}},\ }\href
  {https://doi.org/10.1016/j.nima.2008.11.076} {\bibfield  {journal} {\bibinfo
  {journal} {Nucl. Instrum. Meth. A}\ }\textbf {\bibinfo {volume} {600}},\
  \bibinfo {pages} {568} (\bibinfo {year} {2009})},\ \Eprint
  {https://arxiv.org/abs/0806.2400} {arXiv:0806.2400 [physics.ins-det]}
  \BibitemShut {NoStop}%
\bibitem [{\citenamefont {Djurcic}\ \emph {et~al.}(2015)\citenamefont {Djurcic}
  \emph {et~al.}}]{JUNO:2015sjr}%
  \BibitemOpen
  \bibfield  {author} {\bibinfo {author} {\bibfnamefont {Z.}~\bibnamefont
  {Djurcic}} \emph {et~al.} (\bibinfo {collaboration} {JUNO}),\ }\bibfield
  {title} {\bibinfo {title} {{JUNO Conceptual Design Report}},\ }\href@noop {}
  {\  (\bibinfo {year} {2015})},\ \Eprint {https://arxiv.org/abs/1508.07166}
  {arXiv:1508.07166 [physics.ins-det]} \BibitemShut {NoStop}%
\bibitem [{\citenamefont {Agostini}\ \emph {et~al.}(2019)\citenamefont
  {Agostini} \emph {et~al.}}]{Borexino:2017rsf}%
  \BibitemOpen
  \bibfield  {author} {\bibinfo {author} {\bibfnamefont {M.}~\bibnamefont
  {Agostini}} \emph {et~al.} (\bibinfo {collaboration} {Borexino}),\ }\bibfield
   {title} {\bibinfo {title} {{First Simultaneous Precision Spectroscopy of
  $pp$, $^7$Be, and $pep$ Solar Neutrinos with Borexino Phase-II}},\ }\href
  {https://doi.org/10.1103/PhysRevD.100.082004} {\bibfield  {journal} {\bibinfo
   {journal} {Phys. Rev. D}\ }\textbf {\bibinfo {volume} {100}},\ \bibinfo
  {pages} {082004} (\bibinfo {year} {2019})},\ \Eprint
  {https://arxiv.org/abs/1707.09279} {arXiv:1707.09279 [hep-ex]} \BibitemShut
  {NoStop}%
\bibitem [{\citenamefont {Agostini}\ \emph
  {et~al.}(2020{\natexlab{a}})\citenamefont {Agostini} \emph
  {et~al.}}]{BOREXINO:2020hox}%
  \BibitemOpen
  \bibfield  {author} {\bibinfo {author} {\bibfnamefont {M.}~\bibnamefont
  {Agostini}} \emph {et~al.} (\bibinfo {collaboration} {BOREXINO}),\ }\bibfield
   {title} {\bibinfo {title} {{Sensitivity to neutrinos from the solar CNO
  cycle in Borexino}},\ }\href
  {https://doi.org/10.1140/epjc/s10052-020-08534-2} {\bibfield  {journal}
  {\bibinfo  {journal} {Eur. Phys. J. C}\ }\textbf {\bibinfo {volume} {80}},\
  \bibinfo {pages} {1091} (\bibinfo {year} {2020}{\natexlab{a}})},\ \Eprint
  {https://arxiv.org/abs/2005.12829} {arXiv:2005.12829 [hep-ex]} \BibitemShut
  {NoStop}%
\bibitem [{\citenamefont {Agostini}\ \emph
  {et~al.}(2020{\natexlab{b}})\citenamefont {Agostini} \emph
  {et~al.}}]{BOREXINO:2020aww}%
  \BibitemOpen
  \bibfield  {author} {\bibinfo {author} {\bibfnamefont {M.}~\bibnamefont
  {Agostini}} \emph {et~al.} (\bibinfo {collaboration} {BOREXINO}),\ }\bibfield
   {title} {\bibinfo {title} {{Experimental evidence of neutrinos produced in
  the CNO fusion cycle in the Sun}},\ }\href
  {https://doi.org/10.1038/s41586-020-2934-0} {\bibfield  {journal} {\bibinfo
  {journal} {Nature}\ }\textbf {\bibinfo {volume} {587}},\ \bibinfo {pages}
  {577} (\bibinfo {year} {2020}{\natexlab{b}})},\ \Eprint
  {https://arxiv.org/abs/2006.15115} {arXiv:2006.15115 [hep-ex]} \BibitemShut
  {NoStop}%
\bibitem [{\citenamefont {Coloma}\ \emph {et~al.}(2014)\citenamefont {Coloma},
  \citenamefont {Huber},\ and\ \citenamefont {Link}}]{Coloma:2014hka}%
  \BibitemOpen
  \bibfield  {author} {\bibinfo {author} {\bibfnamefont {P.}~\bibnamefont
  {Coloma}}, \bibinfo {author} {\bibfnamefont {P.}~\bibnamefont {Huber}},\ and\
  \bibinfo {author} {\bibfnamefont {J.~M.}\ \bibnamefont {Link}},\ }\bibfield
  {title} {\bibinfo {title} {{Combining dark matter detectors and
  electron-capture sources to hunt for new physics in the neutrino sector}},\
  }\href {https://doi.org/10.1007/JHEP11(2014)042} {\bibfield  {journal}
  {\bibinfo  {journal} {JHEP}\ }\textbf {\bibinfo {volume} {11}},\ \bibinfo
  {pages} {042}},\ \Eprint {https://arxiv.org/abs/1406.4914} {arXiv:1406.4914
  [hep-ph]} \BibitemShut {NoStop}%
\bibitem [{\citenamefont {Gavrin}\ \emph {et~al.}(2021)\citenamefont {Gavrin},
  \citenamefont {Ibragimova}, \citenamefont {Kozlova}, \citenamefont {Tarasov},
  \citenamefont {Veretenkin},\ and\ \citenamefont {Zvir}}]{Gavrin:2021lam}%
  \BibitemOpen
  \bibfield  {author} {\bibinfo {author} {\bibfnamefont {V.~N.}\ \bibnamefont
  {Gavrin}}, \bibinfo {author} {\bibfnamefont {T.~V.}\ \bibnamefont
  {Ibragimova}}, \bibinfo {author} {\bibfnamefont {J.~P.}\ \bibnamefont
  {Kozlova}}, \bibinfo {author} {\bibfnamefont {V.~A.}\ \bibnamefont
  {Tarasov}}, \bibinfo {author} {\bibfnamefont {E.~P.}\ \bibnamefont
  {Veretenkin}},\ and\ \bibinfo {author} {\bibfnamefont {A.~I.}\ \bibnamefont
  {Zvir}},\ }\bibfield  {title} {\bibinfo {title} {{Measurement of neutrino
  source activity in the experiment BEST by calorimetric method}},\ }\href
  {https://doi.org/10.1088/1748-0221/16/06/E06001} {\bibfield  {journal}
  {\bibinfo  {journal} {JINST}\ }\textbf {\bibinfo {volume} {16}}\bibfield
  {number} {\bibinfo  {number} { (04)},\ \bibinfo {pages} {P04012}},\ }\Eprint
  {https://arxiv.org/abs/2105.05120} {arXiv:2105.05120 [hep-ex]} \BibitemShut
  {NoStop}%
\bibitem [{\citenamefont {Seltzer}(1993)}]{NIST}%
  \BibitemOpen
  \bibfield  {author} {\bibinfo {author} {\bibfnamefont {S.~M.}\ \bibnamefont
  {Seltzer}},\ }\bibfield  {title} {\bibinfo {title} {Calculation of photon
  mass energy-transfer and mass energy-absorption coefficients},\ }\href@noop
  {} {\bibfield  {journal} {\bibinfo  {journal} {Radiation research}\ }\textbf
  {\bibinfo {volume} {136}},\ \bibinfo {pages} {147} (\bibinfo {year}
  {1993})}\BibitemShut {NoStop}%
\bibitem [{\citenamefont {Arpesella}\ \emph {et~al.}(2002)\citenamefont
  {Arpesella} \emph {et~al.}}]{BOREXINO:2001bob}%
  \BibitemOpen
  \bibfield  {author} {\bibinfo {author} {\bibfnamefont {C.}~\bibnamefont
  {Arpesella}} \emph {et~al.} (\bibinfo {collaboration} {BOREXINO}),\
  }\bibfield  {title} {\bibinfo {title} {{Measurements of extremely low
  radioactivity levels in BOREXINO}},\ }\href
  {https://doi.org/10.1016/S0927-6505(01)00179-7} {\bibfield  {journal}
  {\bibinfo  {journal} {Astropart. Phys.}\ }\textbf {\bibinfo {volume} {18}},\
  \bibinfo {pages} {1} (\bibinfo {year} {2002})},\ \Eprint
  {https://arxiv.org/abs/hep-ex/0109031} {arXiv:hep-ex/0109031} \BibitemShut
  {NoStop}%
\bibitem [{\citenamefont {Wang}\ \emph {et~al.}(2002)\citenamefont {Wang},
  \citenamefont {Wong},\ and\ \citenamefont {Fujiwara}}]{WANG2002498}%
  \BibitemOpen
  \bibfield  {author} {\bibinfo {author} {\bibfnamefont {S.}~\bibnamefont
  {Wang}}, \bibinfo {author} {\bibfnamefont {H.}~\bibnamefont {Wong}},\ and\
  \bibinfo {author} {\bibfnamefont {M.}~\bibnamefont {Fujiwara}},\ }\bibfield
  {title} {\bibinfo {title} {Measurement of intrinsic radioactivity in a gso
  crystal},\ }\href
  {https://doi.org/https://doi.org/10.1016/S0168-9002(01)00929-9} {\bibfield
  {journal} {\bibinfo  {journal} {Nuclear Instruments and Methods in Physics
  Research Section A: Accelerators, Spectrometers, Detectors and Associated
  Equipment}\ }\textbf {\bibinfo {volume} {479}},\ \bibinfo {pages} {498}
  (\bibinfo {year} {2002})}\BibitemShut {NoStop}%
\bibitem [{\citenamefont {Li}\ \emph {et~al.}(2001)\citenamefont {Li} \emph
  {et~al.}}]{TEXONO:2000zzq}%
  \BibitemOpen
  \bibfield  {author} {\bibinfo {author} {\bibfnamefont {H.~B.}\ \bibnamefont
  {Li}} \emph {et~al.} (\bibinfo {collaboration} {TEXONO}),\ }\bibfield
  {title} {\bibinfo {title} {{A CsI(Tl) scintillating crystal detector for the
  studies of low-energy neutrino interactions}},\ }\href
  {https://doi.org/10.1016/S0168-9002(00)00999-2} {\bibfield  {journal}
  {\bibinfo  {journal} {Nucl. Instrum. Meth. A}\ }\textbf {\bibinfo {volume}
  {459}},\ \bibinfo {pages} {93} (\bibinfo {year} {2001})},\ \bibinfo {note}
  {[Erratum: Nucl.Instrum.Meth.A 485, 821 (2002)]},\ \Eprint
  {https://arxiv.org/abs/hep-ex/0001001} {arXiv:hep-ex/0001001} \BibitemShut
  {NoStop}%
\bibitem [{\citenamefont {Suerfu}\ \emph {et~al.}(2020)\citenamefont {Suerfu},
  \citenamefont {Wada}, \citenamefont {Peloso}, \citenamefont {Souza},
  \citenamefont {Calaprice}, \citenamefont {Tower},\ and\ \citenamefont
  {Ciampi}}]{PhysRevResearch.2.013223}%
  \BibitemOpen
  \bibfield  {author} {\bibinfo {author} {\bibfnamefont {B.}~\bibnamefont
  {Suerfu}}, \bibinfo {author} {\bibfnamefont {M.}~\bibnamefont {Wada}},
  \bibinfo {author} {\bibfnamefont {W.}~\bibnamefont {Peloso}}, \bibinfo
  {author} {\bibfnamefont {M.}~\bibnamefont {Souza}}, \bibinfo {author}
  {\bibfnamefont {F.}~\bibnamefont {Calaprice}}, \bibinfo {author}
  {\bibfnamefont {J.}~\bibnamefont {Tower}},\ and\ \bibinfo {author}
  {\bibfnamefont {G.}~\bibnamefont {Ciampi}},\ }\bibfield  {title} {\bibinfo
  {title} {Growth of ultra-high purity nai(tl) crystals for dark matter
  searches},\ }\href {https://doi.org/10.1103/PhysRevResearch.2.013223}
  {\bibfield  {journal} {\bibinfo  {journal} {Phys. Rev. Res.}\ }\textbf
  {\bibinfo {volume} {2}},\ \bibinfo {pages} {013223} (\bibinfo {year}
  {2020})}\BibitemShut {NoStop}%
\bibitem [{\citenamefont {Hosokawa}\ \emph {et~al.}(2022)\citenamefont
  {Hosokawa}, \citenamefont {Ikeda}, \citenamefont {Okada}, \citenamefont
  {Sekiya}, \citenamefont {Fernández}, \citenamefont {Labarga}, \citenamefont
  {Bandac}, \citenamefont {Perez}, \citenamefont {Ito}, \citenamefont {Harada},
  \citenamefont {Koshio}, \citenamefont {Thiesse}, \citenamefont {Thompson},
  \citenamefont {Scovell}, \citenamefont {Meehan}, \citenamefont {Ichimura},
  \citenamefont {Kishimoto}, \citenamefont {Nakajima}, \citenamefont {Vagins},
  \citenamefont {Ito}, \citenamefont {Takaku}, \citenamefont {Tanaka},\ and\
  \citenamefont {Yamaguchi}}]{superkgd}%
  \BibitemOpen
  \bibfield  {author} {\bibinfo {author} {\bibfnamefont {K.}~\bibnamefont
  {Hosokawa}}, \bibinfo {author} {\bibfnamefont {M.}~\bibnamefont {Ikeda}},
  \bibinfo {author} {\bibfnamefont {T.}~\bibnamefont {Okada}}, \bibinfo
  {author} {\bibfnamefont {H.}~\bibnamefont {Sekiya}}, \bibinfo {author}
  {\bibfnamefont {P.}~\bibnamefont {Fernández}}, \bibinfo {author}
  {\bibfnamefont {L.}~\bibnamefont {Labarga}}, \bibinfo {author} {\bibfnamefont
  {I.}~\bibnamefont {Bandac}}, \bibinfo {author} {\bibfnamefont
  {J.}~\bibnamefont {Perez}}, \bibinfo {author} {\bibfnamefont
  {S.}~\bibnamefont {Ito}}, \bibinfo {author} {\bibfnamefont {M.}~\bibnamefont
  {Harada}}, \bibinfo {author} {\bibfnamefont {Y.}~\bibnamefont {Koshio}},
  \bibinfo {author} {\bibfnamefont {M.~D.}\ \bibnamefont {Thiesse}}, \bibinfo
  {author} {\bibfnamefont {L.~F.}\ \bibnamefont {Thompson}}, \bibinfo {author}
  {\bibfnamefont {P.~R.}\ \bibnamefont {Scovell}}, \bibinfo {author}
  {\bibfnamefont {E.}~\bibnamefont {Meehan}}, \bibinfo {author} {\bibfnamefont
  {K.}~\bibnamefont {Ichimura}}, \bibinfo {author} {\bibfnamefont
  {Y.}~\bibnamefont {Kishimoto}}, \bibinfo {author} {\bibfnamefont
  {Y.}~\bibnamefont {Nakajima}}, \bibinfo {author} {\bibfnamefont {M.~R.}\
  \bibnamefont {Vagins}}, \bibinfo {author} {\bibfnamefont {H.}~\bibnamefont
  {Ito}}, \bibinfo {author} {\bibfnamefont {Y.}~\bibnamefont {Takaku}},
  \bibinfo {author} {\bibfnamefont {Y.}~\bibnamefont {Tanaka}},\ and\ \bibinfo
  {author} {\bibfnamefont {Y.}~\bibnamefont {Yamaguchi}},\ }\bibfield  {title}
  {\bibinfo {title} {{Development of ultra-pure gadolinium sulfate for the
  Super-Kamiokande gadolinium project}},\ }\bibfield  {journal} {\bibinfo
  {journal} {Progress of Theoretical and Experimental Physics}\ }\textbf
  {\bibinfo {volume} {2023}},\ \href {https://doi.org/10.1093/ptep/ptac170}
  {10.1093/ptep/ptac170} (\bibinfo {year} {2022}),\ \bibinfo {note} {013H01},\
  \Eprint
  {https://arxiv.org/abs/https://academic.oup.com/ptep/article-pdf/2023/1/013H01/48939277/ptac170.pdf}
  {https://academic.oup.com/ptep/article-pdf/2023/1/013H01/48939277/ptac170.pdf}
  \BibitemShut {NoStop}%
\bibitem [{ens()}]{ensdf}%
  \BibitemOpen
  \bibfield  {title} {\bibinfo {title} {\protect{ENSDF} database},\ }\bibinfo
  {note} {accessed 02/06/2023}\BibitemShut {NoStop}%
\bibitem [{\citenamefont {Furuno}\ \emph {et~al.}(2021)\citenamefont {Furuno},
  \citenamefont {Koshikawa}, \citenamefont {Kawabata}, \citenamefont {Itoh},
  \citenamefont {Kurosawa}, \citenamefont {Morimoto}, \citenamefont {Murata},
  \citenamefont {Sakanashi}, \citenamefont {Tsumura},\ and\ \citenamefont
  {Yamaji}}]{Furuno2021}%
  \BibitemOpen
  \bibfield  {author} {\bibinfo {author} {\bibfnamefont {T.}~\bibnamefont
  {Furuno}}, \bibinfo {author} {\bibfnamefont {A.}~\bibnamefont {Koshikawa}},
  \bibinfo {author} {\bibfnamefont {T.}~\bibnamefont {Kawabata}}, \bibinfo
  {author} {\bibfnamefont {M.}~\bibnamefont {Itoh}}, \bibinfo {author}
  {\bibfnamefont {S.}~\bibnamefont {Kurosawa}}, \bibinfo {author}
  {\bibfnamefont {T.}~\bibnamefont {Morimoto}}, \bibinfo {author}
  {\bibfnamefont {M.}~\bibnamefont {Murata}}, \bibinfo {author} {\bibfnamefont
  {K.}~\bibnamefont {Sakanashi}}, \bibinfo {author} {\bibfnamefont
  {M.}~\bibnamefont {Tsumura}},\ and\ \bibinfo {author} {\bibfnamefont
  {A.}~\bibnamefont {Yamaji}},\ }\bibfield  {title} {\bibinfo {title} {Response
  of the gagg(ce) scintillator to charged particles compared with the csi(tl)
  scintillator},\ }\href {https://doi.org/10.1088/1748-0221/16/10/P10012}
  {\bibfield  {journal} {\bibinfo  {journal} {Journal of Instrumentation}\
  }\textbf {\bibinfo {volume} {16}}\bibinfo  {number} { (10)},\ \bibinfo
  {pages} {P10012}}\BibitemShut {NoStop}%
\bibitem [{\citenamefont {{Sibczy{\'n}ski}}\ \emph {et~al.}(2018)\citenamefont
  {{Sibczy{\'n}ski}}, \citenamefont {{Czarnacki}}, \citenamefont {{Mianowska}},
  \citenamefont {{Mianowski}}, \citenamefont {{Moszy{\'n}ski}}, \citenamefont
  {{Sworobowicz}}, \citenamefont {{{\'S}widerski}}, \citenamefont {{Bezbakh}},
  \citenamefont {{Fomichev}}, \citenamefont {{Krupko}}, \citenamefont
  {{Sabelnikov}}, \citenamefont {{Kamada}}, \citenamefont {{Shoji}},\ and\
  \citenamefont {{Yoshikawa}}}]{2018NIMPA}%
  \BibitemOpen
\bibfield  {number} {  }\bibfield  {author} {\bibinfo {author} {\bibfnamefont
  {P.}~\bibnamefont {{Sibczy{\'n}ski}}}, \bibinfo {author} {\bibfnamefont
  {W.}~\bibnamefont {{Czarnacki}}}, \bibinfo {author} {\bibfnamefont
  {Z.}~\bibnamefont {{Mianowska}}}, \bibinfo {author} {\bibfnamefont
  {S.}~\bibnamefont {{Mianowski}}}, \bibinfo {author} {\bibfnamefont
  {M.}~\bibnamefont {{Moszy{\'n}ski}}}, \bibinfo {author} {\bibfnamefont
  {T.}~\bibnamefont {{Sworobowicz}}}, \bibinfo {author} {\bibfnamefont
  {{\L}.}~\bibnamefont {{{\'S}widerski}}}, \bibinfo {author} {\bibfnamefont
  {A.~A.}\ \bibnamefont {{Bezbakh}}}, \bibinfo {author} {\bibfnamefont {A.~S.}\
  \bibnamefont {{Fomichev}}}, \bibinfo {author} {\bibfnamefont {S.~A.}\
  \bibnamefont {{Krupko}}}, \bibinfo {author} {\bibfnamefont {A.~V.}\
  \bibnamefont {{Sabelnikov}}}, \bibinfo {author} {\bibfnamefont
  {K.}~\bibnamefont {{Kamada}}}, \bibinfo {author} {\bibfnamefont
  {Y.}~\bibnamefont {{Shoji}}},\ and\ \bibinfo {author} {\bibfnamefont
  {A.}~\bibnamefont {{Yoshikawa}}},\ }\bibfield  {title} {\bibinfo {title}
  {{Non-proportionality of GAGG:Ce scintillators down to 50 eV electron
  equivalent by application of alpha particle excitation}},\ }\href
  {https://doi.org/10.1016/j.nima.2018.03.050} {\bibfield  {journal} {\bibinfo
  {journal} {Nuclear Instruments and Methods in Physics Research A}\ }\textbf
  {\bibinfo {volume} {898}},\ \bibinfo {pages} {24} (\bibinfo {year}
  {2018})}\BibitemShut {NoStop}%
\bibitem [{\citenamefont {Huber}(2011)}]{Huber:2011wv}%
  \BibitemOpen
  \bibfield  {author} {\bibinfo {author} {\bibfnamefont {P.}~\bibnamefont
  {Huber}},\ }\bibfield  {title} {\bibinfo {title} {{On the determination of
  anti-neutrino spectra from nuclear reactors}},\ }\href
  {https://doi.org/10.1103/PhysRevC.85.029901} {\bibfield  {journal} {\bibinfo
  {journal} {Phys. Rev. C}\ }\textbf {\bibinfo {volume} {84}},\ \bibinfo
  {pages} {024617} (\bibinfo {year} {2011})},\ \bibinfo {note} {[Erratum:
  Phys.Rev.C 85, 029901 (2012)]},\ \Eprint {https://arxiv.org/abs/1106.0687}
  {arXiv:1106.0687 [hep-ph]} \BibitemShut {NoStop}%
\bibitem [{\citenamefont {Amare}\ \emph {et~al.}(2018)\citenamefont {Amare}
  \emph {et~al.}}]{Amare:2017roa}%
  \BibitemOpen
  \bibfield  {author} {\bibinfo {author} {\bibfnamefont {J.}~\bibnamefont
  {Amare}} \emph {et~al.},\ }\bibfield  {title} {\bibinfo {title} {{Cosmogenic
  production of tritium in dark matter detectors}},\ }\href
  {https://doi.org/10.1016/j.astropartphys.2017.11.004} {\bibfield  {journal}
  {\bibinfo  {journal} {Astropart. Phys.}\ }\textbf {\bibinfo {volume} {97}},\
  \bibinfo {pages} {96} (\bibinfo {year} {2018})},\ \Eprint
  {https://arxiv.org/abs/1706.05818} {arXiv:1706.05818 [physics.ins-det]}
  \BibitemShut {NoStop}%
\bibitem [{\citenamefont {Saldanha}\ \emph {et~al.}(2020)\citenamefont
  {Saldanha} \emph {et~al.}}]{Saldanha:2020ubf}%
  \BibitemOpen
  \bibfield  {author} {\bibinfo {author} {\bibfnamefont {R.}~\bibnamefont
  {Saldanha}} \emph {et~al.},\ }\bibfield  {title} {\bibinfo {title}
  {{Cosmogenic activation of silicon}},\ }\href
  {https://doi.org/10.1103/PhysRevD.102.102006} {\bibfield  {journal} {\bibinfo
   {journal} {Phys. Rev. D}\ }\textbf {\bibinfo {volume} {102}},\ \bibinfo
  {pages} {102006} (\bibinfo {year} {2020})},\ \Eprint
  {https://arxiv.org/abs/2007.10584} {arXiv:2007.10584 [physics.ins-det]}
  \BibitemShut {NoStop}%
\bibitem [{\citenamefont {Raghavan}\ \emph {et~al.}(2008)\citenamefont
  {Raghavan}, \citenamefont {collaboration} \emph {et~al.}}]{raghavan2008lens}%
  \BibitemOpen
  \bibfield  {author} {\bibinfo {author} {\bibfnamefont {R.}~\bibnamefont
  {Raghavan}}, \bibinfo {author} {\bibfnamefont {L.}~\bibnamefont
  {collaboration}}, \emph {et~al.},\ }\bibfield  {title} {\bibinfo {title}
  {Lens, minilens—status and outlook},\ }in\ \href@noop {} {\emph {\bibinfo
  {booktitle} {Journal of Physics: Conference Series}}},\ Vol.\ \bibinfo
  {volume} {120}\ (\bibinfo {organization} {IOP Publishing},\ \bibinfo {year}
  {2008})\ p.\ \bibinfo {pages} {052014}\BibitemShut {NoStop}%
\end{thebibliography}%

\end{document}